%% file: candar.tex
\documentclass{amsart}
\usepackage{cite}

\usepackage{color,hyperref}
\usepackage{amsthm}
\usepackage{amssymb}
\usepackage{amsmath}
\usepackage{gnuplot-lua-tikz}
\usepackage{tikz}
\usepackage{tikz-cd}
\usetikzlibrary{matrix,arrows,positioning,automata,shapes}
\tikzset{
   level distance=1.2cm,sibling distance=.8cm,
   edge from parent path={(\tikzparentnode) -- (\tikzchildnode)},
   elb/.style={draw,ellipse,inner sep=1pt}, 
   wlb/.style={fill=white,inner sep=1pt}, 
   glb/.style={fill=black!7,inner sep=1pt}, 
   gelb/.style={draw,ellipse,fill=black!7,inner sep=1pt}, 
   nlb/.style={inner sep=0pt,draw=none,minimum size=0pt}, 
   highlight/.style={line width=.5cm,color=black!22,cap=round,join=round,opacity=0.3}
}
\pgfdeclarelayer{background layer}
\pgfsetlayers{background layer,main}

\usepackage{url}

\theoremstyle{plain}
\newtheorem{theorem}{Theorem}

\theoremstyle{definition}
\newtheorem{definition}[theorem]{Definition}
\newtheorem{example}[theorem]{Example}
\newtheorem{principle}[theorem]{Principle}

\newcommand{\N}{\mathbb N}
\newcommand{\Z}{\mathbb{Z}}

\newcommand{\B}[1]{\textbf{#1}}

\newcommand{\marginnote}[1]{}

\newcommand{\FullTrans}{\mathcal T}
\newcommand{\Symmetric}{\mathcal S}

\newcommand{\Sub}{\mathbf{Sub}}

\begin{document}
\title{Finite Computational Structures and Implementations}
\thanks{The final version of this paper will be presented at CANDAR'16, Fourth International Symposium on Computing and Networking, Hiroshima, Japan, November 22-25, 2016, and it will be published by the IEEE Computer Society.}
\author{Attila Egri-Nagy}
\address{Akita International University, Japan}
\email{\url{egri-nagy@aiu.ac.jp}}
\urladdr{www.egri-nagy.hu}
\maketitle
\begin{abstract}
\emph{What is computable with limited resources?}
\emph{How can we verify the correctness of computations?}
\emph{How to measure computational power with precision?}
Despite the immense scientific and engineering progress in computing, we still have only partial answers to these questions.
In order to make these problems more precise, we describe an abstract algebraic definition of classical computation, generalizing traditional models to semigroups.
The mathematical abstraction also allows the investigation of different computing paradigms (e.g.~cellular automata, reversible computing) in the same framework.
Here we summarize the main questions and recent results of the research of finite computation.
\end{abstract}

\section{Introduction}

The exponential growth of the computing power of hardware (colloquially known as Moore's Law)
seems to be ended by reaching its physical and economical limits.
In order to keep up technological development, producing more mathematical knowledge about digital computation is crucial for improving the efficiency of software.
Complementing research in computational complexity, where the emphasis is on the asymptotic behaviour of algorithms, we need to refocus on small computing devices, and study the possibilities of limited finite computations.

\section{Computational structures}

\emph{What is classical computation?}
Dictionary definitions are somewhat circular, e.g.~computation is what a computer does and a computer is a device that performs computation.
A quick look at actual computing devices reveals that computation is
\begin{enumerate}
\item a mapping from inputs to outputs;
\item a sequence of state transitions;
\item described by mathematical models;
\item implemented by physical systems;
\item a hierarchical structure;
\item potentially universal.
\end{enumerate}
The first two points seem to be opposites \emph{What to compute?} versus \emph{How to compute?},
high versus low-level descriptions, declarative versus imperative programming paradigms.
However, as we will show, they are just the two extremes of the same computation spectrum.

Our computers are physical devices and the theory of computation is abstracted from physical processes.
Mathematical models clearly define the notion of computation, but mapping the abstract computation back to the physical realm is often considered problematic.
We argue that structure-preserving maps between computations work from one mathematical model to another just as well as from the abstract  to the concrete physical implementation, easily crossing any ontological borderline one might assume between the two. The former needs mathematical thinking, the latter engineering, but the underlying problem is the same: find relations between computing structures that do not change the computed function.
Since abstract algebra provides the required tools, we suggest further abstractions to the models of computations to reach the algebraic level safe for discussing implementations.
It is also suitable for capturing the hierarchical structure of computers.
Finiteness and the abstract algebraic approach paint a picture where universal computation becomes relative and the `mathematical versus physical' distinction less important.

First we attempt to define computations and implementations purely as abstract functions, then the need for combining functions leads us to definition of computational structures.

\subsection{Computation as a function: input-output mappings}
Starting from the human perspective, computation is a tool.
We want a solution for some problem: the input is the question, the output is the answer.
Formally, the input-output pairs are represented as a \emph{function} $f:X\rightarrow Y$, and computation is function evaluation $f(x)=y$, $x\in X, y\in Y$.
 As an implementation of $f$, we need a dynamical system whose behaviour  can be modelled by another function $g$, which is essentially the same as $f$.
\begin{center}
\begin{tikzcd}
X \arrow[r,"f"] \arrow[d,"\varphi_1"] & Y \arrow[d,"\varphi_2"] \\
A \arrow[r,"g"] & B
\end{tikzcd}
\end{center}
We have three ingredients, the function $f$ (specification), the dynamical system $g$ (in the sense of initial conditions and laws of motions, whenever $a\in A$ is present then $g$ will naturally take it to $b=g(a)$), and the pair of mappings $\varphi_1,\varphi_2$ establishing the implementation relation.
As in the abstract theory of functions (category theory, see for instance \cite{awodey2006category}), the above diagram commutes: $\varphi_2\left(f(x)\right)=g(\varphi_1(x))$.
\marginnote{\begin{example}Let $f:\N \rightarrow \N$ be the function $f(x)=x+2$. Let $\varphi_1(x)=\varphi_2(x)$ be the action of raising $x$ many fingers, mapping zero to closed fists. Let $g$ be the action of raising two un-raised fingers. Then $g$ is a working implementation of $f$ for $x\leq 8$.\end{example}}
This means that if the value of the abstract function is $y=f(x)$ at $x$, then after finding the `physical' representation of $x$, namely $\varphi_1(x)$, the dynamics of $g$ will produce $g(\varphi_1(x))$. This has to be the same as the physical representation of $y$, namely $\varphi_2(f(x))$.
In order to make the implementation useful, we require $\varphi_1, \varphi_2$ to be one-to-one. This ensures, that $f$ and $g$ are the same, up to some relabelling. Therefore, properties of the functions are also preserved. For instance, an invertible function can only be implemented by another invertible function, and vice versa.
In the theory of computable functions we start from a set of primitive functions and build composite functions by combining them \cite{cutland1980computability}.
Thus, \emph{function composition} is a fundamental way of constructing computations.

Getting closer to real computations, we need to fill the elements of the abstract sets for input and output with some content.
The content can be represented by bitstrings.
Thus, computation can be described by a mapping  $$f:\{0,1\}^m\rightarrow \{0,1\}^n,\quad m,n\in\N.$$
According to the fundamental theorem of reversible computing, any finite function can be computed by an invertible function \cite{toffoli1980reversible}.
This apparently contradicts the idea of implementation, that important properties of functions have to be preserved.
There seems to be a way to sidestep function isomorphism.
\begin{example}\label{ex:xor}
    Embedding \textsc{XOR}
\begin{center}
    \begin{tabular}{ccc}
      \colorbox{gray!30}{00} &$\mapsto$&\colorbox{gray!30}{0}0\\
      \colorbox{gray!30}{01}&$\mapsto$&\colorbox{gray!30}{1}1\\
      \colorbox{gray!30}{10}&$\mapsto$&\colorbox{gray!30}{1}0\\
      \colorbox{gray!30}{11}&$\mapsto$&\colorbox{gray!30}{0}1\\
    \end{tabular}
    \end{center}
    \noindent and \textsc{FAN-OUT}
\begin{center}
    \begin{tabular}{ccc}
      0\colorbox{gray!30}{0} &$\mapsto$&\colorbox{gray!30}{00}\\
      0\colorbox{gray!30}{1}&$\mapsto$&\colorbox{gray!30}{11}\\
      10&$\mapsto$&10\\
      11&$\mapsto$&01\\
    \end{tabular}
    \end{center}
\noindent into the same bijective function. By putting information into the abstract elements, any function can `piggyback' even on the identity function.
\end{example}
Another example of this sidestepping is generating pseudo-random numbers, producing randomness from a non-random deterministic process. One method involves multiplying big numbers and cutting some digits out from the middle.
These `tricks' work by composing the actual computations with special input and output functions, that might have different properties.
In reversible computing the readout operation may not be a reversible function.

\subsection{Computation as a process: state transitions}
Focusing on the process view, what is the most basic unit of computation? A \emph{state transition}: an event changes the current state of a system.
A \emph{state} is defined by a configuration of a system's components, or some other distinctive properties the system may have. In classical computing, the assumption is that the states are well-defined and easily distinguishable, discrete entities. In analog computing state varies along a continuum, while in quantum computing states are in superposition.

Let's say the current state is $x$, then event $s$ happens changing the state to $y$. We might write $y=s(x)$ emphasizing that $s$ is a function, but it is better to write
$$ xs=y$$
meaning that  $s$ happens in state $x$ yielding state $y$. Why?
Because combining events as they happen one after the other, e.g.~$xst=z$, is easier to read following our left to right convention.

Though it is more intuitive to distinguish between states and events, it is not a fundamental distinction.

\begin{principle}[State-event abstraction]
\marginnote{In abstract algebra, this is known as the Cayley's theorem for semigroups \cite{Howie95}.}
We can identify an event with its resulting state: state $x$ is where we end up when event $x$ happens.
\end{principle}
\noindent  According to the \emph{action interpretation}, $xs=y$ can be understood as event $s$ changes the current state $x$ to the next state $y$.
But $xs=y$ can also be read as event $x$ combined with event $s$ yields the composite event $y$, the \emph{event algebra interpretation}.

We can combine abstract events into longer sequences.
These can also be considered as a sequence of instructions, i.e.~an \emph{algorithm} \cite{tslang2010}.
These sequences of events should have the property of associativity
$$(xy)z=x(yz) \text{ for all abstract events } x,y,z,$$
since a given sequence $xyz$ should be a well-defined algorithm.
This also shows that we can reason about algorithms using equations.

We can put all event combinations into a table. These are the rules describing how to combine any two events.
\begin{definition}[Computational Structure]
A finite set $X$ and a rule for combining elements of $X$ that assigns a value $x'$ for each two-element sequence, written as $xy=x'$, is a \emph{computational structure} if $(xy)z=x(yz)$ for all $x,y,z\in X$.
\end{definition}
In mathematics, a set closed under an associative  binary operation is an abstract algebraic structure called \emph{semigroup}. The somewhat derogative term is used because of historical reasons. Group theory advanced first, so semigroups are considered as broken groups, and not the other way around, groups as (very important) special cases of semigroups.
\begin{example}[Flip-flop, 1-bit memory semigroup]
\begin{center}\begin{tabular}{c|ccc}
 &$r$&0&1\\
\hline
$r$&$r$&0&1\\
0&0&0&1\\
1&1&0&1
\end{tabular}
\end{center}
\noindent The read-operation is $r$. Events $0$ and $1$ correspond to destructive storage of bit values. Algebraically these are \emph{right zero elements}, or simply \emph{resets}.
\end{example}

Computation is a process in time  -- an obvious assumption, since most of engineering and complexity studies are about doing computation in shorter time.
Combining two events yield a third one (which can be the same), and we can continue with combining them to have an ordered sequence of events.  This ordering may be referred as time.
However, at the abstraction level of the state transition table time is not essential.
The table implicitly describes all possible sequences of events, it defines the rules how to combine any two events, but it is a timeless entity. This is similar to some controversial ideas in theoretical physics \cite{barbour2001end}.



\subsection{The computation spectrum}

How are the function and the process view of computation related? They are actually the same.
Given a computable function, we can construct a computational structure capable of computing the function. An algorithm  (a sequence of state transition events) takes an initial state (encoded input) into a final state (encoded output).
The simplest way to achieve this is by a lookup table.
\begin{definition}[Lookup table semigroup]
Let $f:X\rightarrow Y$ be a function, where $X\cap Y=\varnothing$. Then the semigroup $S=X\cup Y\cup \{\ell\}$ consists of resets $X\cup Y$ and the lookup operation $\ell$ defined by $x\ell=y$ if $f(x)=y$ for all $x\in X$ and $u\ell=u$ for all $u\in S\setminus X$.
\end{definition}
\noindent Is it associative? Let $v\in X\cup Y$ be an arbitrary reset element, and $s,t\in S$ any element.
Since the rightmost event is a reset, we have $(st)v=v$ and $s(tv)=sv=v$.
For $(sv)\ell=v\ell=s(v\ell)$ since $v\ell$ is also a reset.
For $(v\ell)\ell=v\ell$, since $\ell$ does not change anything in $S\setminus X$ and  $v(\ell\ell)=v\ell$ since $\ell$ is an idempotent ($\ell\ell=\ell$).
Separating the domain and the codomain of $f$ is crucial, for functions $X\rightarrow X$ we can simply have  a separate copy of elements of $X$.
When trying to make it more economical associativity may not be easy to achieve \cite{1978assocundec}.

Turning a computational structure into a function is also easy. Pick any algorithm (a composite event), and that is also a function from states to states.

Why do we have different approaches then? Different computations can realize the same function. In software engineering, for optimization purposes we often use pre-calculated data, and just look the value up in a table when it is needed, thus saving time. Another technique computes the value on demand, but stores it for later queries (caching, memoization). This observation motivates the following questions.
\emph{How much processing is done in a computation? How many state transitions?} Based on this we have a whole spectrum of computation, from mere storage and retrieval to computations producing data from little input.
Information storage and retrieval are forms of computation.
By the same token computation can be considered as a general form of information storage and retrieval, where looking up the required piece of data may need many steps.
We can say that if computation is information processing, then information is frozen computation.
For instance, when calculating logarithms were slow and difficult, one had to use tabulated values in a printed book.
Combinatorial (stateless) circuits are another example of lookup table computations.
Arithmetic calculations (by humans) are somewhere in the middle of the spectrum. We add and multiply single digit numbers with lookup table method, but using a cascade algorithm for longer numbers.
The other extreme consists of computations that do not rely on any input data, though they may fill the memory up for later usage.
For instance, busy beaver Turing-machines \cite{BusyBeaver1995}, or the hypothetical shortest program that generates the sequence of bitstrings describing the consecutive states in the evolution of our observable universe \cite{Schmidhuber1997}.

\subsection{Traditional mathematical models of computation}

This algebraic approach may look different from the more mainstream models of computation \cite{savage1998models}, algebraic automata theory \cite{holcombe_textbook} being a sub-field of theoretical computer science.
However, considering computational structures as semigroups is a stronger abstraction, therefore the algebraic treatment of computations is actually more general and thus more fundamental.

From finite state automata, we abstract away the initial and accepting states.
Those special states are needed only for the important application of recognizing languages.
Input symbols of a finite state automaton are fully defined transformations (total functions) of its state set.
\begin{definition}
A \emph{transformation} is a function $f:X\rightarrow X$ from a set to itself,
and a \emph{transformation semigroup} $(X,S)$ of degree $n$ is a collection $S$ of transformations of an $n$-element set closed under function composition.
\end{definition}
\noindent If we focus on the possible state transitions of a finite state automaton only, we get a transformation semigroup with a generator set corresponding to the input symbols. These semigroups are very special representations of abstract semigroups, where state transition is realized by composing functions.

A Turing-machine without the infinite tape is also a finite state automaton.
A finite length tape can always be incorporated into the state set of the automaton.
In general, if we take those models of computation that describe the detailed dynamics of computation, and remove all the model specific decorations, we get a semigroup.

\subsection{Computers: physical realizations of computation}

Intuitively, a computer is a physical system whose dynamics at some level can be described as a computational structure.
For any equation $xy=z$ in the computational structure, we should be able to induce in the physical system an event corresponding to $x$ and another one corresponding to $y$ such that their overall effect corresponds to $z$.
Algebraically, this special correspondence is a structure-preserving map, a \emph{homomorphism}.
If we want exact realizations, not just approximations, then we need stricter one-to-one mappings, \emph{isomorphisms}.
However, for computational structures we need to use relations instead of functions.

\begin{definition}[Isomorphic relations of computational structures]
  Let $S$ and $T$ be computational structures (semigroups).
  A  relation $\varphi:S\rightarrow T$ is an \emph{isomorphic relation} if it is
\begin{enumerate}
\item homomorphic: $\varphi(s)\varphi(t)\subseteq \varphi(st)$,
\item fully defined: $\varphi(s)\neq \varnothing$ for all $s \in S$,
\item lossless: $\varphi(s)\cap\varphi(t)\neq\varnothing\implies s=t$
\end{enumerate}
for all $s,t\in S$. We also say that $T$ \emph{emulates}, or \emph{implements} $S$.
\end{definition}
\emph{Homomorphic} is the key property, it ensures that similar computation is done in $T$ by matching individual state transitions.
Here $\varphi(s)$ and $\varphi(t)$ are subsets of $T$ (not just single elements), and $\varphi(s)\varphi(t)$ denotes all possible state transitions induced by these subsets (element-wise product of two sets).
 \emph{Fully defined} means that we assign some state(s) of $T$ for all elements of $S$, so we account for everything that can happen in $S$.
In general, homomorphic maps are structure-forgetting, since we can map several states to a single one. Being \emph{lossless} excludes loosing information about state transitions.
In semigroup theory, isomorphic relations are called \emph{divisions}, a special type of \emph{relational morphisms} \cite{QBook}.

What happens if we turn an implementation around? It becomes a computational model.
\begin{definition}[Modelling of computational structures]
\marginnote{In mathematics this is called \emph{surjective homomorphism}}
  Let $S$ and $T$ be computational structures (semigroups).
  A function $\mu:T\rightarrow S$ is a \emph{modelling} if it is
\begin{enumerate}
\item homomorphism: $\mu(u)\mu(v)= \mu(uv)$ for all $u,v\in T$,
\item onto: for all $s \in S$ there exists a $u\in T$ such that $\mu(u)=s$.
\end{enumerate}
We also say that $S$ is a \emph{computational model} of $T$. In algebra, functions of this kind are called \emph{surjective homomorphisms}.
\end{definition}
A modelling is a function, so it is fully defined. A modelling $\mu$ turned around $\mu^{-1}$ is an implementation, and an implementation $\varphi$ turned around is a modelling $\varphi^{-1}$.
This is an asymmetric relation, naturally we assume that a model of any system is smaller in some sense than the system itself. Also, to implement a computational structure completely we need another structure at least as big.

According to the mathematical universe hypothesis\cite{TMU2008}, we have nothing more to do, since we covered mappings from one mathematical structure to another one.
In practice, we do seem to have a distinction between mathematical models of computations and actual computers, since abstract models by definition are abstracted away from reality, they do not have any inherent dynamical force to carry out state transitions. Even pen and paper calculations require a driving force, the human hand and the pattern matching capabilities of the brain.
But we can apply a simple strategy: we treat a physical system as a mathematical structure, regardless its ontological status. Building a computer then becomes the task of constructing an isomorphic relation.
\begin{definition}[Computer]
An implementation of a computational structure by a physical system is a \emph{computer}.
\end{definition}
Finding such a relation to a physical system is highly non-trivial. Charles Babbage failed to establish the correspondence between arithmetical operations and the mechanisms of cogwheels.
However, the key point is to see that it is just another mapping. Maybe more difficult for a physical implementation, but it is not different from establishing emulation relation between abstract computational models.

Anything that is capable of state transitions can be used for some computation. The question is how useful that computation is?
We can always map the target system's mathematical model onto itself. In this sense the cosmos can be considered as a giant computer computing itself. However, this statement is a bit hollow since we do not have a complete mathematical model of the universe.
\marginnote{``In a sense, nature has been continually computing the `next state' of the universe for billions of years; all we have to do -- and, actually, all we can do -- is `hitch a ride' on this huge ongoing computation , and try to discover which parts of it happen to go near to where we want.'' \cite{toffoli1982physics}}
Every physical system computes, at least its future states, but not everything does useful calculation. \marginnote{\emph{Is the universe a giant computer?} The universe only computes its next state, so we can only map its mathematical model to it, and we do not have its complete description. Moreover, since by definition it takes no input from outside, its dynamics is disappointingly simple (monogenic semigroup). However, it is interesting that we have parts of it capable of universal computation.} Much like entropy is heat not available for useful work.
The same way as steam and combustion engines exploit physical processes to process materials and move things around, computers exploit physical processes to transfer and transform information.

\subsection{Hierarchical structure}
\begin{figure}
\tikzset{->,>=triangle 45,auto,node distance=2cm}
\begin{tikzpicture}
\tikzstyle{every state}=[minimum size=3pt]
  \node[state]  (q_a)  {$00$};
  \node[state]  (q_b) [right of=q_a] {$11$};
  \node[state]  (q_c) [right of=q_b] {$01$};
  \node[state]  (q_d) [right of=q_c] {$10$};
  \node[fill=black!10]  (r_0) [below of=q_b] {$r_0$};
  \node[state]  (q_0) [below of=r_0] {$0$};
  \node[fill=black!10]  (r_1) [below of=q_c] {$r_1$};
  \node[state]  (q_1) [below of=r_1] {$1$};

  \path (q_a) edge  [loop] node [above] {$t$} (q_a);
  \path (q_d) edge  [loop] node [above] {$t$} (q_a);
   \path (q_b) edge  [bend left] node [above] {$t$} (q_c);
   \path (q_c) edge  [bend left] node [above] {$t$} (q_b);

  \path (q_0) edge  [loop above] node (r0) [above] {$$} (q_0);
   \path (q_1) edge  [bend right] node (r0_) [above] {$$} (q_0);
   \path (q_0) edge  [bend right] node [above] (r1_) {$$} (q_1);
  \path (q_1) edge  [loop above] node (r1) [above] {$$} (q_1);
  \path[>=angle 60,densely dotted]  (r_0) edge (r0);
  \path[>=angle 60,densely dotted]  (r_0) edge (r0_);
  \path[>=angle 60,densely dotted]  (r_1) edge (r1);
  \path[>=angle 60,densely dotted]  (r_1) edge (r1_);

   \path[>=angle 60,densely dotted]  (q_a) edge (r_0);
   \path[>=angle 60,densely dotted]  (q_b) edge (r_0);
   \path[>=angle 60,densely dotted]  (q_c) edge (r_1);
   \path[>=angle 60,densely dotted]  (q_d) edge (r_1);
\end{tikzpicture}
\caption{Hierarchical, transformation semigroup construction of reversible \textsc{XOR} function (see Example \ref{ex:xor}).
We combine two transformation semigroups into a composite one. The independent component (4 states at the top) is a reversible permutation group with state transition $t$ defining the dynamics, while the dependent component (2 states at the bottom, the readout part) has reset operations $r_0, r_1$. State transitions in the bottom component are chosen based on the state of the top component. These are indicated by the dashed line. If the top level component is in state 00 or 11, then the bottom component resets to $0$ by state transition $r_0$. Otherwise, $r_1$ is carried out. The input is set in the top level component (by choosing a state), and the output is the resulting state in the bottom component.}
\label{fig:XORcascade}
\end{figure}
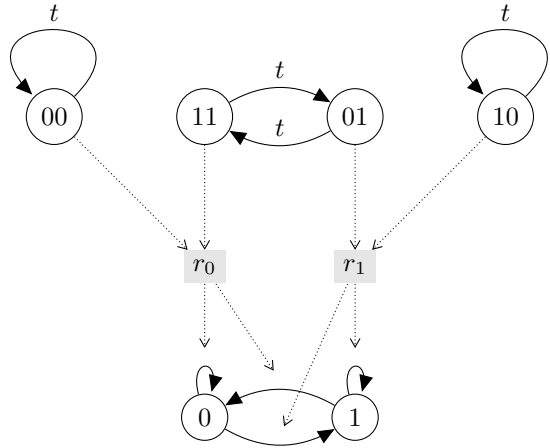

Huge state transition tables are not particularly useful to look at; they are like  quark-level descriptions for trying to understand living organisms.
Identifying substructures and their interactions is needed.
Hierarchical levels of organizations provide an important way to  understand computers.
Information flow is limited to one-way only along a partial order, thus enabling functional abstraction.
\begin{figure*}
\begin{center}
\begin{tabular}{|c|c|c|}
\hline
 &  \B{Natural Numbers} & \B{Computational Structures}  \\
\hline
\B{Building Blocks} & Primes & 1-bit memory\\
           &         &  reversible computations \\
\hline
\B{Composition} & Multiplication & Cascade Product \\
\hline
\B{Precision}  & Equality & Emulation \\
\hline
\B{Uniqueness}  & Unique & Non-unique \\
\hline
\end{tabular}
\end{center}
\caption{A comparison of the prime decomposition of integers and computational structures. Cascade decompositions are ways of understanding systems modeled by finite state automata. This type of modeling is used increasingly for biological systems (e.g.~\cite{WeakControl2015}).}
\label{fig:decomp}
\end{figure*}
According to Krohn-Rhodes theory \cite{primedecomp68}, any computational structure can be built by using destructive memory storage and the reversible computational structures in a hierarchical manner (Fig.~\ref{fig:decomp}).
The way the components are put together is the \emph{cascade product} \cite{cascprod}, which is a substructure of the algebraic wreath product.
The distinction between reversible and irreversible is sharp: there is no way to embed state collapsing into a permutation structure.
Reversible computing \cite{toffoli1980reversible} seems to contradict this. The trick there is to put information into  states and then selectively read off partial information from the resulting states.
This selection of required information can be done by another computational structure.
We can have a reversible computational structure on the top, and one at the bottom that implements the readout. We can have many state transitions in the reversible part without a readout (Fig.~\ref{fig:XORcascade}).
Reversible implementations may prove to be decisive in terms of power efficiency of computers, but it does not erase the difference between reversible and irreversible computations.

Important to note that hierarchical decompositions are even possible when the computational structure is not hierarchical itself. One of the research directions is the study of how  it is possible to understand loopback systems in a hierarchical manner.
Fortunately, now we have computer algebra tools available for generating these decompositions \cite{SgpDec2014}.
\subsection{Universal computers}

\emph{What is the difference between a piece of rock and a silicon chip?}
They are made of the same material, but they have different computational power. The rock only computes its next state (its temperature, all the wiggling of its atoms), so the only computation we can map to it homomorphically is its own mathematical description. While the silicon chip admits other computational structures. General purpose processors are homomorphic images of universal computers.

Universality is a fact of everyday computing (programs run other programs, computers emulate other type of computers).
It is also a central concept in computability theory \cite{Turing1936,AnnotatedTuring2008}.
The universal Turing machine $U$ takes a program $P$, i.e.~a dedicated computer, and the input $x$ of the program. Then by recreating each step of $P$ it computes the result of $P$ on $x$. Formally, $U(P,x)=P(x)$.
Unfortunately, this only makes sense for Turing machines with infinite tape.
With finite resources universal becomes relative, i.e.~universal relative to some kind of representation and size.
A universal semigroup for size $n$ abstract semigroups would be the semigroup that can implement all size $n$ semigroups. There is a trivial construction, a huge direct product of everything  of size maximum $n$. What are the minimal examples of these? --  that is indeed an interesting mathematical question.
For a concrete representation it is easier to find relatively universal structures. For instance, the \emph{full transformation semigroup of degree $n$} (denoted by $\FullTrans_n$) consists of all $n^n$ transformations of an $n$-element set \cite{ClassicalTransSemigroups2009}.

\section{Open problems}

We use computers more and more for extending our knowledge in many scientific fields.
Therefore we need to learn more about the tools as well.
The main topics where further research needs to be done are

\begin{enumerate}
\item exploring the space of possible computations;

\item measuring finite computational power;

\item computational correctness.
\end{enumerate}

\subsection{What are the possible  computational structures and implementations?}

Cataloguing, stocktaking are basic human activities for answering the question \emph{What do we have exactly?}
For the classification of computational structures and implementations, we need to explore the space of all computational structures and their implementations, starting from the small examples. Looking at those is the same as asking the questions \emph{What can we compute with limited resources?}
\emph{What is computable with $n$ states?} This is a complementary approach to computational complexity, where the emphasis is on the growth rate of resource requirements.

\begin{figure*}
\begin{center}
\input{T4size6peaks.tikz}
\end{center}
\caption{The main bulk of the size distribution of transformation semigroups of degree 4. Currently we have no explanation for the six peaks of the distribution, or any information about the asymptotic behaviour of the distribution when the number of states increases.} 
\label{fig:T4size6peaks}
\end{figure*}
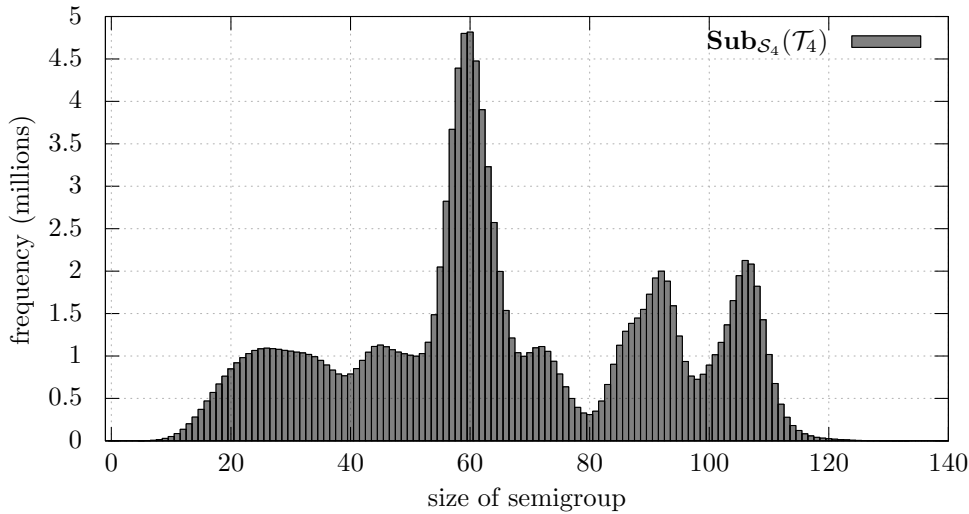

Due to the effect of combinatorial explosion, an exhaustive enumeration of computational structures is doomed to fail eventually. But we need to produce raw data to think about, so we have to push the boundaries of exploration in order to formulate and prove general mathematical results.
Strategy is the following: take a relatively universal structure and enumerate all of its substructures.
For example, finding all transformation semigroups on $n$ states is the same as finding all subsemigroups of $\FullTrans_n$. This strategy lead to the successful enumeration of all 132 069 776 transformation semigroups with 4 states \cite{T4enum} (Fig.~\ref{fig:T4size6peaks}).
This method can be applied to a wider class of semigroups, called diagram semigroups \cite{diagsgps}. These can be considered as `unconventional' mathematical models of computations (e.g.~computing with binary relations or partitions instead of functions).

\subsection{How to measure the amount of computational power?}

\emph{Given an abstract or physical computer, what computations can it perform?} 
The algebraic description gives a clear definition of \emph{emulation}, when one computer can do the job of some other computer. This is a crude form of measuring computational power, in the sense of the `at least as much as' relation.
This way computational power can be measured on a partial order (the lattice of all finite semigroups).

The remaining problem is to bring some computation into the common denominator semigroup form. For example, if we have a finite piece of cellular automata (CA) grid, what can we calculate with it? If the CA is universal and big enough we might be able to fit in a universal Turing machine that would do the required computation. However, we might be able to run my computation directly on the CA instead of a bulky construct. Like given a desktop computer, there is a choice between a high-level interpreted and slow programming language, and the machine-code level. Here, we are interested in native computation.
For cellular automata it is not obvious how to measure the computation. There are different ways to think about it.
\begin{enumerate}
\item Fixed initial conditions: We start the CA from a fixed configuration. The only event is the clock-tick. Algebraically this structure is way too simple. This also resonates with constructor theory, which says that initial conditions and laws of motion have restricted explanatory power \cite{Deutsch2013}.
\item I/O mappings: Take all runs from initial configurations and record the (class of) final configurations. Lookup table style computation.
\item Interaction, perturbation: The events of the modelling automaton are small changes in the grid, acting on stable states or cycles.
\item Piggybacking: Similar to the piggybacking trick of reversible computation, we can use some patch of the CA for input and another (possibly) overlapping patch for the output. Physical universality \cite{physuniversal2014} is also defined this way.
\end{enumerate}

Extending the slogan, ``Numbers measure size, groups measure symmetry.'' \cite{armstrong1988groups}, we can say that semigroups measure computation.
\subsection{How can we trust computers?}

Even in mathematics, we increasingly rely on creating knowledge by computers.
It is impractical to fully check the isomorphic relation between the computational structure and the implementing physical system.
We do not actually have the complete computational structure, only a set of generators.
In a way computation can be viewed as generating computational structures from a partial description.
But how can we be sure that the relation works for combined operations?
A physical system does more, not just the mapped computation. How can we make sure that nothing leaks into the abstract computation from the underlying physics?
The actual paths of computation, the sequences  of events may interact more than what is described in the abstract state transitions.


What can we do?
In practice, we do test hardware and software, but not all possible computations.
We build up confidence by solving the same problem repeatedly by using several different methods and many different computers. Compare for instance \textsc{SgpDec}~\cite{SgpDec} and \textsc{Kigen}~\cite{kigen}.
This practical approach raises theoretical questions.
\emph{When are two solutions really different?} Computations can differ by
\begin{enumerate}
\item having different intermediate results,
\item applying different operations,
\item having different modular structure,
\end{enumerate}
and by any combination of these.

The first two are in accord with the problem of enumeration and classification of computational structures and implementations: find all distinct, minimal computational processes realizing the same function. In other words, within a universal computational structure, what are the different ways of realizing the same computational structure?
Current research attacks this problem by finding efficient algorithms for constructing implementations (embeddings) of arbitrary abstract computational structures \cite{repcon}.

\section{Conclusion}

We suggest that generalizing existing models of computation to semigroup theory will help in solving open problems in software and hardware engineering. In turn, the mathematical investigation relies on the tools of high-performance computing, forming a positive feedback loop.
Therefore, despite the current gap between the practical computing problems and the scale of the exact mathematical results, the systematic study of finite computation is bound to produce groundbreaking results.
\bibliography{comp}
\bibliographystyle{plain}
\end{document}

%% file: T4size6peaks.tikz
\begin{tikzpicture}[gnuplot]
\path (1.000,0.000) rectangle (14.000,7.000);
\gpcolor{color=gp lt color axes}
\gpsetlinetype{gp lt axes}
\gpsetlinewidth{1.00}
\draw[gp path] (2.240,0.985)--(13.447,0.985);
\gpcolor{color=gp lt color border}
\gpsetlinetype{gp lt border}
\draw[gp path] (2.240,0.985)--(2.420,0.985);
\draw[gp path] (13.447,0.985)--(13.267,0.985);
\node[gp node right] at (2.056,0.985) { 0};
\gpcolor{color=gp lt color axes}
\gpsetlinetype{gp lt axes}
\draw[gp path] (2.240,1.550)--(13.447,1.550);
\gpcolor{color=gp lt color border}
\gpsetlinetype{gp lt border}
\draw[gp path] (2.240,1.550)--(2.420,1.550);
\draw[gp path] (13.447,1.550)--(13.267,1.550);
\node[gp node right] at (2.056,1.550) { 0.5};
\gpcolor{color=gp lt color axes}
\gpsetlinetype{gp lt axes}
\draw[gp path] (2.240,2.114)--(13.447,2.114);
\gpcolor{color=gp lt color border}
\gpsetlinetype{gp lt border}
\draw[gp path] (2.240,2.114)--(2.420,2.114);
\draw[gp path] (13.447,2.114)--(13.267,2.114);
\node[gp node right] at (2.056,2.114) { 1};
\gpcolor{color=gp lt color axes}
\gpsetlinetype{gp lt axes}
\draw[gp path] (2.240,2.679)--(13.447,2.679);
\gpcolor{color=gp lt color border}
\gpsetlinetype{gp lt border}
\draw[gp path] (2.240,2.679)--(2.420,2.679);
\draw[gp path] (13.447,2.679)--(13.267,2.679);
\node[gp node right] at (2.056,2.679) { 1.5};
\gpcolor{color=gp lt color axes}
\gpsetlinetype{gp lt axes}
\draw[gp path] (2.240,3.243)--(13.447,3.243);
\gpcolor{color=gp lt color border}
\gpsetlinetype{gp lt border}
\draw[gp path] (2.240,3.243)--(2.420,3.243);
\draw[gp path] (13.447,3.243)--(13.267,3.243);
\node[gp node right] at (2.056,3.243) { 2};
\gpcolor{color=gp lt color axes}
\gpsetlinetype{gp lt axes}
\draw[gp path] (2.240,3.808)--(13.447,3.808);
\gpcolor{color=gp lt color border}
\gpsetlinetype{gp lt border}
\draw[gp path] (2.240,3.808)--(2.420,3.808);
\draw[gp path] (13.447,3.808)--(13.267,3.808);
\node[gp node right] at (2.056,3.808) { 2.5};
\gpcolor{color=gp lt color axes}
\gpsetlinetype{gp lt axes}
\draw[gp path] (2.240,4.373)--(13.447,4.373);
\gpcolor{color=gp lt color border}
\gpsetlinetype{gp lt border}
\draw[gp path] (2.240,4.373)--(2.420,4.373);
\draw[gp path] (13.447,4.373)--(13.267,4.373);
\node[gp node right] at (2.056,4.373) { 3};
\gpcolor{color=gp lt color axes}
\gpsetlinetype{gp lt axes}
\draw[gp path] (2.240,4.937)--(13.447,4.937);
\gpcolor{color=gp lt color border}
\gpsetlinetype{gp lt border}
\draw[gp path] (2.240,4.937)--(2.420,4.937);
\draw[gp path] (13.447,4.937)--(13.267,4.937);
\node[gp node right] at (2.056,4.937) { 3.5};
\gpcolor{color=gp lt color axes}
\gpsetlinetype{gp lt axes}
\draw[gp path] (2.240,5.502)--(13.447,5.502);
\gpcolor{color=gp lt color border}
\gpsetlinetype{gp lt border}
\draw[gp path] (2.240,5.502)--(2.420,5.502);
\draw[gp path] (13.447,5.502)--(13.267,5.502);
\node[gp node right] at (2.056,5.502) { 4};
\gpcolor{color=gp lt color axes}
\gpsetlinetype{gp lt axes}
\draw[gp path] (2.240,6.066)--(13.447,6.066);
\gpcolor{color=gp lt color border}
\gpsetlinetype{gp lt border}
\draw[gp path] (2.240,6.066)--(2.420,6.066);
\draw[gp path] (13.447,6.066)--(13.267,6.066);
\node[gp node right] at (2.056,6.066) { 4.5};
\gpcolor{color=gp lt color axes}
\gpsetlinetype{gp lt axes}
\draw[gp path] (2.240,6.631)--(13.447,6.631);
\gpcolor{color=gp lt color border}
\gpsetlinetype{gp lt border}
\draw[gp path] (2.240,6.631)--(2.420,6.631);
\draw[gp path] (13.447,6.631)--(13.267,6.631);
\node[gp node right] at (2.056,6.631) { 5 };
\gpcolor{color=gp lt color axes}
\gpsetlinetype{gp lt axes}
\draw[gp path] (2.319,0.985)--(2.319,6.631);
\gpcolor{color=gp lt color border}
\gpsetlinetype{gp lt border}
\draw[gp path] (2.319,0.985)--(2.319,1.165);
\draw[gp path] (2.319,6.631)--(2.319,6.451);
\node[gp node center] at (2.319,0.677) { 0};
\gpcolor{color=gp lt color axes}
\gpsetlinetype{gp lt axes}
\draw[gp path] (3.909,0.985)--(3.909,6.631);
\gpcolor{color=gp lt color border}
\gpsetlinetype{gp lt border}
\draw[gp path] (3.909,0.985)--(3.909,1.165);
\draw[gp path] (3.909,6.631)--(3.909,6.451);
\node[gp node center] at (3.909,0.677) { 20};
\gpcolor{color=gp lt color axes}
\gpsetlinetype{gp lt axes}
\draw[gp path] (5.499,0.985)--(5.499,6.631);
\gpcolor{color=gp lt color border}
\gpsetlinetype{gp lt border}
\draw[gp path] (5.499,0.985)--(5.499,1.165);
\draw[gp path] (5.499,6.631)--(5.499,6.451);
\node[gp node center] at (5.499,0.677) { 40};
\gpcolor{color=gp lt color axes}
\gpsetlinetype{gp lt axes}
\draw[gp path] (7.088,0.985)--(7.088,6.631);
\gpcolor{color=gp lt color border}
\gpsetlinetype{gp lt border}
\draw[gp path] (7.088,0.985)--(7.088,1.165);
\draw[gp path] (7.088,6.631)--(7.088,6.451);
\node[gp node center] at (7.088,0.677) { 60};
\gpcolor{color=gp lt color axes}
\gpsetlinetype{gp lt axes}
\draw[gp path] (8.678,0.985)--(8.678,6.631);
\gpcolor{color=gp lt color border}
\gpsetlinetype{gp lt border}
\draw[gp path] (8.678,0.985)--(8.678,1.165);
\draw[gp path] (8.678,6.631)--(8.678,6.451);
\node[gp node center] at (8.678,0.677) { 80};
\gpcolor{color=gp lt color axes}
\gpsetlinetype{gp lt axes}
\draw[gp path] (10.268,0.985)--(10.268,6.631);
\gpcolor{color=gp lt color border}
\gpsetlinetype{gp lt border}
\draw[gp path] (10.268,0.985)--(10.268,1.165);
\draw[gp path] (10.268,6.631)--(10.268,6.451);
\node[gp node center] at (10.268,0.677) { 100};
\gpcolor{color=gp lt color axes}
\gpsetlinetype{gp lt axes}
\draw[gp path] (11.857,0.985)--(11.857,6.143);
\draw[gp path] (11.857,6.451)--(11.857,6.631);
\gpcolor{color=gp lt color border}
\gpsetlinetype{gp lt border}
\draw[gp path] (11.857,0.985)--(11.857,1.165);
\draw[gp path] (11.857,6.631)--(11.857,6.451);
\node[gp node center] at (11.857,0.677) { 120};
\gpcolor{color=gp lt color axes}
\gpsetlinetype{gp lt axes}
\draw[gp path] (13.447,0.985)--(13.447,6.631);
\gpcolor{color=gp lt color border}
\gpsetlinetype{gp lt border}
\draw[gp path] (13.447,0.985)--(13.447,1.165);
\draw[gp path] (13.447,6.631)--(13.447,6.451);
\node[gp node center] at (13.447,0.677) { 140};
\draw[gp path] (2.240,6.631)--(2.240,0.985)--(13.447,0.985)--(13.447,6.631)--cycle;
\node[gp node center,rotate=-270] at (1.146,3.808) {frequency (millions)};
\node[gp node center] at (7.843,0.215) {size of semigroup};
\node[gp node right] at (11.979,6.297) {$\Sub_{\Symmetric_4}(\FullTrans_4)$};
\gpfill{rgb color={0.000,0.000,0.000},opacity=0.50} (12.163,6.220)--(13.079,6.220)--(13.079,6.374)--(12.163,6.374)--cycle;
\gpcolor{rgb color={0.000,0.000,0.000}}
\gpsetlinetype{gp lt plot 0}
\draw[gp path] (12.163,6.220)--(13.079,6.220)--(13.079,6.374)--(12.163,6.374)--cycle;
\gpfill{rgb color={0.000,0.000,0.000},opacity=0.50} (2.280,0.985)--(2.360,0.985)--(2.360,0.986)--(2.280,0.986)--cycle;
\draw[gp path] (2.280,0.985)--(2.359,0.985)--cycle;
\gpfill{rgb color={0.000,0.000,0.000},opacity=0.50} (2.359,0.985)--(2.440,0.985)--(2.440,0.986)--(2.359,0.986)--cycle;
\draw[gp path] (2.359,0.985)--(2.439,0.985)--cycle;
\gpfill{rgb color={0.000,0.000,0.000},opacity=0.50} (2.439,0.985)--(2.519,0.985)--(2.519,0.986)--(2.439,0.986)--cycle;
\draw[gp path] (2.439,0.985)--(2.518,0.985)--cycle;
\gpfill{rgb color={0.000,0.000,0.000},opacity=0.50} (2.518,0.985)--(2.599,0.985)--(2.599,0.986)--(2.518,0.986)--cycle;
\draw[gp path] (2.518,0.985)--(2.598,0.985)--cycle;
\gpfill{rgb color={0.000,0.000,0.000},opacity=0.50} (2.598,0.985)--(2.678,0.985)--(2.678,0.986)--(2.598,0.986)--cycle;
\draw[gp path] (2.598,0.985)--(2.677,0.985)--cycle;
\gpfill{rgb color={0.000,0.000,0.000},opacity=0.50} (2.677,0.985)--(2.758,0.985)--(2.758,0.987)--(2.677,0.987)--cycle;
\draw[gp path] (2.677,0.985)--(2.677,0.986)--(2.757,0.986)--(2.757,0.985)--cycle;
\gpfill{rgb color={0.000,0.000,0.000},opacity=0.50} (2.757,0.985)--(2.837,0.985)--(2.837,0.989)--(2.757,0.989)--cycle;
\draw[gp path] (2.757,0.985)--(2.757,0.988)--(2.836,0.988)--(2.836,0.985)--cycle;
\gpfill{rgb color={0.000,0.000,0.000},opacity=0.50} (2.836,0.985)--(2.917,0.985)--(2.917,0.994)--(2.836,0.994)--cycle;
\draw[gp path] (2.836,0.985)--(2.836,0.993)--(2.916,0.993)--(2.916,0.985)--cycle;
\gpfill{rgb color={0.000,0.000,0.000},opacity=0.50} (2.916,0.985)--(2.996,0.985)--(2.996,1.003)--(2.916,1.003)--cycle;
\draw[gp path] (2.916,0.985)--(2.916,1.002)--(2.995,1.002)--(2.995,0.985)--cycle;
\gpfill{rgb color={0.000,0.000,0.000},opacity=0.50} (2.995,0.985)--(3.076,0.985)--(3.076,1.019)--(2.995,1.019)--cycle;
\draw[gp path] (2.995,0.985)--(2.995,1.018)--(3.075,1.018)--(3.075,0.985)--cycle;
\gpfill{rgb color={0.000,0.000,0.000},opacity=0.50} (3.075,0.985)--(3.155,0.985)--(3.155,1.044)--(3.075,1.044)--cycle;
\draw[gp path] (3.075,0.985)--(3.075,1.043)--(3.154,1.043)--(3.154,0.985)--cycle;
\gpfill{rgb color={0.000,0.000,0.000},opacity=0.50} (3.154,0.985)--(3.235,0.985)--(3.235,1.083)--(3.154,1.083)--cycle;
\draw[gp path] (3.154,0.985)--(3.154,1.082)--(3.234,1.082)--(3.234,0.985)--cycle;
\gpfill{rgb color={0.000,0.000,0.000},opacity=0.50} (3.234,0.985)--(3.314,0.985)--(3.314,1.140)--(3.234,1.140)--cycle;
\draw[gp path] (3.234,0.985)--(3.234,1.139)--(3.313,1.139)--(3.313,0.985)--cycle;
\gpfill{rgb color={0.000,0.000,0.000},opacity=0.50} (3.313,0.985)--(3.393,0.985)--(3.393,1.213)--(3.313,1.213)--cycle;
\draw[gp path] (3.313,0.985)--(3.313,1.212)--(3.392,1.212)--(3.392,0.985)--cycle;
\gpfill{rgb color={0.000,0.000,0.000},opacity=0.50} (3.392,0.985)--(3.473,0.985)--(3.473,1.303)--(3.392,1.303)--cycle;
\draw[gp path] (3.392,0.985)--(3.392,1.302)--(3.472,1.302)--(3.472,0.985)--cycle;
\gpfill{rgb color={0.000,0.000,0.000},opacity=0.50} (3.472,0.985)--(3.552,0.985)--(3.552,1.405)--(3.472,1.405)--cycle;
\draw[gp path] (3.472,0.985)--(3.472,1.404)--(3.551,1.404)--(3.551,0.985)--cycle;
\gpfill{rgb color={0.000,0.000,0.000},opacity=0.50} (3.551,0.985)--(3.632,0.985)--(3.632,1.517)--(3.551,1.517)--cycle;
\draw[gp path] (3.551,0.985)--(3.551,1.516)--(3.631,1.516)--(3.631,0.985)--cycle;
\gpfill{rgb color={0.000,0.000,0.000},opacity=0.50} (3.631,0.985)--(3.711,0.985)--(3.711,1.630)--(3.631,1.630)--cycle;
\draw[gp path] (3.631,0.985)--(3.631,1.629)--(3.710,1.629)--(3.710,0.985)--cycle;
\gpfill{rgb color={0.000,0.000,0.000},opacity=0.50} (3.710,0.985)--(3.791,0.985)--(3.791,1.743)--(3.710,1.743)--cycle;
\draw[gp path] (3.710,0.985)--(3.710,1.742)--(3.790,1.742)--(3.790,0.985)--cycle;
\gpfill{rgb color={0.000,0.000,0.000},opacity=0.50} (3.790,0.985)--(3.870,0.985)--(3.870,1.847)--(3.790,1.847)--cycle;
\draw[gp path] (3.790,0.985)--(3.790,1.846)--(3.869,1.846)--(3.869,0.985)--cycle;
\gpfill{rgb color={0.000,0.000,0.000},opacity=0.50} (3.869,0.985)--(3.950,0.985)--(3.950,1.943)--(3.869,1.943)--cycle;
\draw[gp path] (3.869,0.985)--(3.869,1.942)--(3.949,1.942)--(3.949,0.985)--cycle;
\gpfill{rgb color={0.000,0.000,0.000},opacity=0.50} (3.949,0.985)--(4.029,0.985)--(4.029,2.024)--(3.949,2.024)--cycle;
\draw[gp path] (3.949,0.985)--(3.949,2.023)--(4.028,2.023)--(4.028,0.985)--cycle;
\gpfill{rgb color={0.000,0.000,0.000},opacity=0.50} (4.028,0.985)--(4.109,0.985)--(4.109,2.093)--(4.028,2.093)--cycle;
\draw[gp path] (4.028,0.985)--(4.028,2.092)--(4.108,2.092)--(4.108,0.985)--cycle;
\gpfill{rgb color={0.000,0.000,0.000},opacity=0.50} (4.108,0.985)--(4.188,0.985)--(4.188,2.147)--(4.108,2.147)--cycle;
\draw[gp path] (4.108,0.985)--(4.108,2.146)--(4.187,2.146)--(4.187,0.985)--cycle;
\gpfill{rgb color={0.000,0.000,0.000},opacity=0.50} (4.187,0.985)--(4.268,0.985)--(4.268,2.189)--(4.187,2.189)--cycle;
\draw[gp path] (4.187,0.985)--(4.187,2.188)--(4.267,2.188)--(4.267,0.985)--cycle;
\gpfill{rgb color={0.000,0.000,0.000},opacity=0.50} (4.267,0.985)--(4.347,0.985)--(4.347,2.211)--(4.267,2.211)--cycle;
\draw[gp path] (4.267,0.985)--(4.267,2.210)--(4.346,2.210)--(4.346,0.985)--cycle;
\gpfill{rgb color={0.000,0.000,0.000},opacity=0.50} (4.346,0.985)--(4.427,0.985)--(4.427,2.220)--(4.346,2.220)--cycle;
\draw[gp path] (4.346,0.985)--(4.346,2.219)--(4.426,2.219)--(4.426,0.985)--cycle;
\gpfill{rgb color={0.000,0.000,0.000},opacity=0.50} (4.426,0.985)--(4.506,0.985)--(4.506,2.213)--(4.426,2.213)--cycle;
\draw[gp path] (4.426,0.985)--(4.426,2.212)--(4.505,2.212)--(4.505,0.985)--cycle;
\gpfill{rgb color={0.000,0.000,0.000},opacity=0.50} (4.505,0.985)--(4.586,0.985)--(4.586,2.204)--(4.505,2.204)--cycle;
\draw[gp path] (4.505,0.985)--(4.505,2.203)--(4.585,2.203)--(4.585,0.985)--cycle;
\gpfill{rgb color={0.000,0.000,0.000},opacity=0.50} (4.585,0.985)--(4.665,0.985)--(4.665,2.190)--(4.585,2.190)--cycle;
\draw[gp path] (4.585,0.985)--(4.585,2.189)--(4.664,2.189)--(4.664,0.985)--cycle;
\gpfill{rgb color={0.000,0.000,0.000},opacity=0.50} (4.664,0.985)--(4.745,0.985)--(4.745,2.180)--(4.664,2.180)--cycle;
\draw[gp path] (4.664,0.985)--(4.664,2.179)--(4.744,2.179)--(4.744,0.985)--cycle;
\gpfill{rgb color={0.000,0.000,0.000},opacity=0.50} (4.744,0.985)--(4.824,0.985)--(4.824,2.167)--(4.744,2.167)--cycle;
\draw[gp path] (4.744,0.985)--(4.744,2.166)--(4.823,2.166)--(4.823,0.985)--cycle;
\gpfill{rgb color={0.000,0.000,0.000},opacity=0.50} (4.823,0.985)--(4.904,0.985)--(4.904,2.156)--(4.823,2.156)--cycle;
\draw[gp path] (4.823,0.985)--(4.823,2.155)--(4.903,2.155)--(4.903,0.985)--cycle;
\gpfill{rgb color={0.000,0.000,0.000},opacity=0.50} (4.903,0.985)--(4.983,0.985)--(4.983,2.136)--(4.903,2.136)--cycle;
\draw[gp path] (4.903,0.985)--(4.903,2.135)--(4.982,2.135)--(4.982,0.985)--cycle;
\gpfill{rgb color={0.000,0.000,0.000},opacity=0.50} (4.982,0.985)--(5.063,0.985)--(5.063,2.106)--(4.982,2.106)--cycle;
\draw[gp path] (4.982,0.985)--(4.982,2.105)--(5.062,2.105)--(5.062,0.985)--cycle;
\gpfill{rgb color={0.000,0.000,0.000},opacity=0.50} (5.062,0.985)--(5.142,0.985)--(5.142,2.057)--(5.062,2.057)--cycle;
\draw[gp path] (5.062,0.985)--(5.062,2.056)--(5.141,2.056)--(5.141,0.985)--cycle;
\gpfill{rgb color={0.000,0.000,0.000},opacity=0.50} (5.141,0.985)--(5.222,0.985)--(5.222,1.996)--(5.141,1.996)--cycle;
\draw[gp path] (5.141,0.985)--(5.141,1.995)--(5.221,1.995)--(5.221,0.985)--cycle;
\gpfill{rgb color={0.000,0.000,0.000},opacity=0.50} (5.221,0.985)--(5.301,0.985)--(5.301,1.928)--(5.221,1.928)--cycle;
\draw[gp path] (5.221,0.985)--(5.221,1.927)--(5.300,1.927)--(5.300,0.985)--cycle;
\gpfill{rgb color={0.000,0.000,0.000},opacity=0.50} (5.300,0.985)--(5.381,0.985)--(5.381,1.876)--(5.300,1.876)--cycle;
\draw[gp path] (5.300,0.985)--(5.300,1.875)--(5.380,1.875)--(5.380,0.985)--cycle;
\gpfill{rgb color={0.000,0.000,0.000},opacity=0.50} (5.380,0.985)--(5.460,0.985)--(5.460,1.852)--(5.380,1.852)--cycle;
\draw[gp path] (5.380,0.985)--(5.380,1.851)--(5.459,1.851)--(5.459,0.985)--cycle;
\gpfill{rgb color={0.000,0.000,0.000},opacity=0.50} (5.459,0.985)--(5.540,0.985)--(5.540,1.876)--(5.459,1.876)--cycle;
\draw[gp path] (5.459,0.985)--(5.459,1.875)--(5.539,1.875)--(5.539,0.985)--cycle;
\gpfill{rgb color={0.000,0.000,0.000},opacity=0.50} (5.539,0.985)--(5.619,0.985)--(5.619,1.947)--(5.539,1.947)--cycle;
\draw[gp path] (5.539,0.985)--(5.539,1.946)--(5.618,1.946)--(5.618,0.985)--cycle;
\gpfill{rgb color={0.000,0.000,0.000},opacity=0.50} (5.618,0.985)--(5.698,0.985)--(5.698,2.057)--(5.618,2.057)--cycle;
\draw[gp path] (5.618,0.985)--(5.618,2.056)--(5.697,2.056)--(5.697,0.985)--cycle;
\gpfill{rgb color={0.000,0.000,0.000},opacity=0.50} (5.697,0.985)--(5.778,0.985)--(5.778,2.166)--(5.697,2.166)--cycle;
\draw[gp path] (5.697,0.985)--(5.697,2.165)--(5.777,2.165)--(5.777,0.985)--cycle;
\gpfill{rgb color={0.000,0.000,0.000},opacity=0.50} (5.777,0.985)--(5.857,0.985)--(5.857,2.242)--(5.777,2.242)--cycle;
\draw[gp path] (5.777,0.985)--(5.777,2.241)--(5.856,2.241)--(5.856,0.985)--cycle;
\gpfill{rgb color={0.000,0.000,0.000},opacity=0.50} (5.856,0.985)--(5.937,0.985)--(5.937,2.261)--(5.856,2.261)--cycle;
\draw[gp path] (5.856,0.985)--(5.856,2.260)--(5.936,2.260)--(5.936,0.985)--cycle;
\gpfill{rgb color={0.000,0.000,0.000},opacity=0.50} (5.936,0.985)--(6.016,0.985)--(6.016,2.238)--(5.936,2.238)--cycle;
\draw[gp path] (5.936,0.985)--(5.936,2.237)--(6.015,2.237)--(6.015,0.985)--cycle;
\gpfill{rgb color={0.000,0.000,0.000},opacity=0.50} (6.015,0.985)--(6.096,0.985)--(6.096,2.199)--(6.015,2.199)--cycle;
\draw[gp path] (6.015,0.985)--(6.015,2.198)--(6.095,2.198)--(6.095,0.985)--cycle;
\gpfill{rgb color={0.000,0.000,0.000},opacity=0.50} (6.095,0.985)--(6.175,0.985)--(6.175,2.168)--(6.095,2.168)--cycle;
\draw[gp path] (6.095,0.985)--(6.095,2.167)--(6.174,2.167)--(6.174,0.985)--cycle;
\gpfill{rgb color={0.000,0.000,0.000},opacity=0.50} (6.174,0.985)--(6.255,0.985)--(6.255,2.145)--(6.174,2.145)--cycle;
\draw[gp path] (6.174,0.985)--(6.174,2.144)--(6.254,2.144)--(6.254,0.985)--cycle;
\gpfill{rgb color={0.000,0.000,0.000},opacity=0.50} (6.254,0.985)--(6.334,0.985)--(6.334,2.126)--(6.254,2.126)--cycle;
\draw[gp path] (6.254,0.985)--(6.254,2.125)--(6.333,2.125)--(6.333,0.985)--cycle;
\gpfill{rgb color={0.000,0.000,0.000},opacity=0.50} (6.333,0.985)--(6.414,0.985)--(6.414,2.113)--(6.333,2.113)--cycle;
\draw[gp path] (6.333,0.985)--(6.333,2.112)--(6.413,2.112)--(6.413,0.985)--cycle;
\gpfill{rgb color={0.000,0.000,0.000},opacity=0.50} (6.413,0.985)--(6.493,0.985)--(6.493,2.147)--(6.413,2.147)--cycle;
\draw[gp path] (6.413,0.985)--(6.413,2.146)--(6.492,2.146)--(6.492,0.985)--cycle;
\gpfill{rgb color={0.000,0.000,0.000},opacity=0.50} (6.492,0.985)--(6.573,0.985)--(6.573,2.298)--(6.492,2.298)--cycle;
\draw[gp path] (6.492,0.985)--(6.492,2.297)--(6.572,2.297)--(6.572,0.985)--cycle;
\gpfill{rgb color={0.000,0.000,0.000},opacity=0.50} (6.572,0.985)--(6.652,0.985)--(6.652,2.664)--(6.572,2.664)--cycle;
\draw[gp path] (6.572,0.985)--(6.572,2.663)--(6.651,2.663)--(6.651,0.985)--cycle;
\gpfill{rgb color={0.000,0.000,0.000},opacity=0.50} (6.651,0.985)--(6.732,0.985)--(6.732,3.299)--(6.651,3.299)--cycle;
\draw[gp path] (6.651,0.985)--(6.651,3.298)--(6.731,3.298)--(6.731,0.985)--cycle;
\gpfill{rgb color={0.000,0.000,0.000},opacity=0.50} (6.731,0.985)--(6.811,0.985)--(6.811,4.174)--(6.731,4.174)--cycle;
\draw[gp path] (6.731,0.985)--(6.731,4.173)--(6.810,4.173)--(6.810,0.985)--cycle;
\gpfill{rgb color={0.000,0.000,0.000},opacity=0.50} (6.810,0.985)--(6.891,0.985)--(6.891,5.131)--(6.810,5.131)--cycle;
\draw[gp path] (6.810,0.985)--(6.810,5.130)--(6.890,5.130)--(6.890,0.985)--cycle;
\gpfill{rgb color={0.000,0.000,0.000},opacity=0.50} (6.890,0.985)--(6.970,0.985)--(6.970,5.945)--(6.890,5.945)--cycle;
\draw[gp path] (6.890,0.985)--(6.890,5.944)--(6.969,5.944)--(6.969,0.985)--cycle;
\gpfill{rgb color={0.000,0.000,0.000},opacity=0.50} (6.969,0.985)--(7.050,0.985)--(7.050,6.407)--(6.969,6.407)--cycle;
\draw[gp path] (6.969,0.985)--(6.969,6.406)--(7.049,6.406)--(7.049,0.985)--cycle;
\gpfill{rgb color={0.000,0.000,0.000},opacity=0.50} (7.049,0.985)--(7.129,0.985)--(7.129,6.425)--(7.049,6.425)--cycle;
\draw[gp path] (7.049,0.985)--(7.049,6.424)--(7.128,6.424)--(7.128,0.985)--cycle;
\gpfill{rgb color={0.000,0.000,0.000},opacity=0.50} (7.128,0.985)--(7.209,0.985)--(7.209,6.040)--(7.128,6.040)--cycle;
\draw[gp path] (7.128,0.985)--(7.128,6.039)--(7.208,6.039)--(7.208,0.985)--cycle;
\gpfill{rgb color={0.000,0.000,0.000},opacity=0.50} (7.208,0.985)--(7.288,0.985)--(7.288,5.392)--(7.208,5.392)--cycle;
\draw[gp path] (7.208,0.985)--(7.208,5.391)--(7.287,5.391)--(7.287,0.985)--cycle;
\gpfill{rgb color={0.000,0.000,0.000},opacity=0.50} (7.287,0.985)--(7.368,0.985)--(7.368,4.633)--(7.287,4.633)--cycle;
\draw[gp path] (7.287,0.985)--(7.287,4.632)--(7.367,4.632)--(7.367,0.985)--cycle;
\gpfill{rgb color={0.000,0.000,0.000},opacity=0.50} (7.367,0.985)--(7.447,0.985)--(7.447,3.891)--(7.367,3.891)--cycle;
\draw[gp path] (7.367,0.985)--(7.367,3.890)--(7.446,3.890)--(7.446,0.985)--cycle;
\gpfill{rgb color={0.000,0.000,0.000},opacity=0.50} (7.446,0.985)--(7.527,0.985)--(7.527,3.240)--(7.446,3.240)--cycle;
\draw[gp path] (7.446,0.985)--(7.446,3.239)--(7.526,3.239)--(7.526,0.985)--cycle;
\gpfill{rgb color={0.000,0.000,0.000},opacity=0.50} (7.526,0.985)--(7.606,0.985)--(7.606,2.721)--(7.526,2.721)--cycle;
\draw[gp path] (7.526,0.985)--(7.526,2.720)--(7.605,2.720)--(7.605,0.985)--cycle;
\gpfill{rgb color={0.000,0.000,0.000},opacity=0.50} (7.605,0.985)--(7.686,0.985)--(7.686,2.355)--(7.605,2.355)--cycle;
\draw[gp path] (7.605,0.985)--(7.605,2.354)--(7.685,2.354)--(7.685,0.985)--cycle;
\gpfill{rgb color={0.000,0.000,0.000},opacity=0.50} (7.685,0.985)--(7.765,0.985)--(7.765,2.159)--(7.685,2.159)--cycle;
\draw[gp path] (7.685,0.985)--(7.685,2.158)--(7.764,2.158)--(7.764,0.985)--cycle;
\gpfill{rgb color={0.000,0.000,0.000},opacity=0.50} (7.764,0.985)--(7.844,0.985)--(7.844,2.111)--(7.764,2.111)--cycle;
\draw[gp path] (7.764,0.985)--(7.764,2.110)--(7.843,2.110)--(7.843,0.985)--cycle;
\gpfill{rgb color={0.000,0.000,0.000},opacity=0.50} (7.843,0.985)--(7.924,0.985)--(7.924,2.159)--(7.843,2.159)--cycle;
\draw[gp path] (7.843,0.985)--(7.843,2.158)--(7.923,2.158)--(7.923,0.985)--cycle;
\gpfill{rgb color={0.000,0.000,0.000},opacity=0.50} (7.923,0.985)--(8.003,0.985)--(8.003,2.224)--(7.923,2.224)--cycle;
\draw[gp path] (7.923,0.985)--(7.923,2.223)--(8.002,2.223)--(8.002,0.985)--cycle;
\gpfill{rgb color={0.000,0.000,0.000},opacity=0.50} (8.002,0.985)--(8.083,0.985)--(8.083,2.241)--(8.002,2.241)--cycle;
\draw[gp path] (8.002,0.985)--(8.002,2.240)--(8.082,2.240)--(8.082,0.985)--cycle;
\gpfill{rgb color={0.000,0.000,0.000},opacity=0.50} (8.082,0.985)--(8.162,0.985)--(8.162,2.177)--(8.082,2.177)--cycle;
\draw[gp path] (8.082,0.985)--(8.082,2.176)--(8.161,2.176)--(8.161,0.985)--cycle;
\gpfill{rgb color={0.000,0.000,0.000},opacity=0.50} (8.161,0.985)--(8.242,0.985)--(8.242,2.045)--(8.161,2.045)--cycle;
\draw[gp path] (8.161,0.985)--(8.161,2.044)--(8.241,2.044)--(8.241,0.985)--cycle;
\gpfill{rgb color={0.000,0.000,0.000},opacity=0.50} (8.241,0.985)--(8.321,0.985)--(8.321,1.876)--(8.241,1.876)--cycle;
\draw[gp path] (8.241,0.985)--(8.241,1.875)--(8.320,1.875)--(8.320,0.985)--cycle;
\gpfill{rgb color={0.000,0.000,0.000},opacity=0.50} (8.320,0.985)--(8.401,0.985)--(8.401,1.705)--(8.320,1.705)--cycle;
\draw[gp path] (8.320,0.985)--(8.320,1.704)--(8.400,1.704)--(8.400,0.985)--cycle;
\gpfill{rgb color={0.000,0.000,0.000},opacity=0.50} (8.400,0.985)--(8.480,0.985)--(8.480,1.550)--(8.400,1.550)--cycle;
\draw[gp path] (8.400,0.985)--(8.400,1.549)--(8.479,1.549)--(8.479,0.985)--cycle;
\gpfill{rgb color={0.000,0.000,0.000},opacity=0.50} (8.479,0.985)--(8.560,0.985)--(8.560,1.432)--(8.479,1.432)--cycle;
\draw[gp path] (8.479,0.985)--(8.479,1.431)--(8.559,1.431)--(8.559,0.985)--cycle;
\gpfill{rgb color={0.000,0.000,0.000},opacity=0.50} (8.559,0.985)--(8.639,0.985)--(8.639,1.357)--(8.559,1.357)--cycle;
\draw[gp path] (8.559,0.985)--(8.559,1.356)--(8.638,1.356)--(8.638,0.985)--cycle;
\gpfill{rgb color={0.000,0.000,0.000},opacity=0.50} (8.638,0.985)--(8.719,0.985)--(8.719,1.336)--(8.638,1.336)--cycle;
\draw[gp path] (8.638,0.985)--(8.638,1.335)--(8.718,1.335)--(8.718,0.985)--cycle;
\gpfill{rgb color={0.000,0.000,0.000},opacity=0.50} (8.718,0.985)--(8.798,0.985)--(8.798,1.381)--(8.718,1.381)--cycle;
\draw[gp path] (8.718,0.985)--(8.718,1.380)--(8.797,1.380)--(8.797,0.985)--cycle;
\gpfill{rgb color={0.000,0.000,0.000},opacity=0.50} (8.797,0.985)--(8.878,0.985)--(8.878,1.516)--(8.797,1.516)--cycle;
\draw[gp path] (8.797,0.985)--(8.797,1.515)--(8.877,1.515)--(8.877,0.985)--cycle;
\gpfill{rgb color={0.000,0.000,0.000},opacity=0.50} (8.877,0.985)--(8.957,0.985)--(8.957,1.736)--(8.877,1.736)--cycle;
\draw[gp path] (8.877,0.985)--(8.877,1.735)--(8.956,1.735)--(8.956,0.985)--cycle;
\gpfill{rgb color={0.000,0.000,0.000},opacity=0.50} (8.956,0.985)--(9.037,0.985)--(9.037,2.005)--(8.956,2.005)--cycle;
\draw[gp path] (8.956,0.985)--(8.956,2.004)--(9.036,2.004)--(9.036,0.985)--cycle;
\gpfill{rgb color={0.000,0.000,0.000},opacity=0.50} (9.036,0.985)--(9.116,0.985)--(9.116,2.258)--(9.036,2.258)--cycle;
\draw[gp path] (9.036,0.985)--(9.036,2.257)--(9.115,2.257)--(9.115,0.985)--cycle;
\gpfill{rgb color={0.000,0.000,0.000},opacity=0.50} (9.115,0.985)--(9.196,0.985)--(9.196,2.444)--(9.115,2.444)--cycle;
\draw[gp path] (9.115,0.985)--(9.115,2.443)--(9.195,2.443)--(9.195,0.985)--cycle;
\gpfill{rgb color={0.000,0.000,0.000},opacity=0.50} (9.195,0.985)--(9.275,0.985)--(9.275,2.549)--(9.195,2.549)--cycle;
\draw[gp path] (9.195,0.985)--(9.195,2.548)--(9.274,2.548)--(9.274,0.985)--cycle;
\gpfill{rgb color={0.000,0.000,0.000},opacity=0.50} (9.274,0.985)--(9.355,0.985)--(9.355,2.620)--(9.274,2.620)--cycle;
\draw[gp path] (9.274,0.985)--(9.274,2.619)--(9.354,2.619)--(9.354,0.985)--cycle;
\gpfill{rgb color={0.000,0.000,0.000},opacity=0.50} (9.354,0.985)--(9.434,0.985)--(9.434,2.735)--(9.354,2.735)--cycle;
\draw[gp path] (9.354,0.985)--(9.354,2.734)--(9.433,2.734)--(9.433,0.985)--cycle;
\gpfill{rgb color={0.000,0.000,0.000},opacity=0.50} (9.433,0.985)--(9.514,0.985)--(9.514,2.937)--(9.433,2.937)--cycle;
\draw[gp path] (9.433,0.985)--(9.433,2.936)--(9.513,2.936)--(9.513,0.985)--cycle;
\gpfill{rgb color={0.000,0.000,0.000},opacity=0.50} (9.513,0.985)--(9.593,0.985)--(9.593,3.153)--(9.513,3.153)--cycle;
\draw[gp path] (9.513,0.985)--(9.513,3.152)--(9.592,3.152)--(9.592,0.985)--cycle;
\gpfill{rgb color={0.000,0.000,0.000},opacity=0.50} (9.592,0.985)--(9.673,0.985)--(9.673,3.244)--(9.592,3.244)--cycle;
\draw[gp path] (9.592,0.985)--(9.592,3.243)--(9.672,3.243)--(9.672,0.985)--cycle;
\gpfill{rgb color={0.000,0.000,0.000},opacity=0.50} (9.672,0.985)--(9.752,0.985)--(9.752,3.111)--(9.672,3.111)--cycle;
\draw[gp path] (9.672,0.985)--(9.672,3.110)--(9.751,3.110)--(9.751,0.985)--cycle;
\gpfill{rgb color={0.000,0.000,0.000},opacity=0.50} (9.751,0.985)--(9.832,0.985)--(9.832,2.783)--(9.751,2.783)--cycle;
\draw[gp path] (9.751,0.985)--(9.751,2.782)--(9.831,2.782)--(9.831,0.985)--cycle;
\gpfill{rgb color={0.000,0.000,0.000},opacity=0.50} (9.831,0.985)--(9.911,0.985)--(9.911,2.381)--(9.831,2.381)--cycle;
\draw[gp path] (9.831,0.985)--(9.831,2.380)--(9.910,2.380)--(9.910,0.985)--cycle;
\gpfill{rgb color={0.000,0.000,0.000},opacity=0.50} (9.910,0.985)--(9.991,0.985)--(9.991,2.042)--(9.910,2.042)--cycle;
\draw[gp path] (9.910,0.985)--(9.910,2.041)--(9.990,2.041)--(9.990,0.985)--cycle;
\gpfill{rgb color={0.000,0.000,0.000},opacity=0.50} (9.990,0.985)--(10.070,0.985)--(10.070,1.848)--(9.990,1.848)--cycle;
\draw[gp path] (9.990,0.985)--(9.990,1.847)--(10.069,1.847)--(10.069,0.985)--cycle;
\gpfill{rgb color={0.000,0.000,0.000},opacity=0.50} (10.069,0.985)--(10.149,0.985)--(10.149,1.805)--(10.069,1.805)--cycle;
\draw[gp path] (10.069,0.985)--(10.069,1.804)--(10.148,1.804)--(10.148,0.985)--cycle;
\gpfill{rgb color={0.000,0.000,0.000},opacity=0.50} (10.148,0.985)--(10.229,0.985)--(10.229,1.872)--(10.148,1.872)--cycle;
\draw[gp path] (10.148,0.985)--(10.148,1.871)--(10.228,1.871)--(10.228,0.985)--cycle;
\gpfill{rgb color={0.000,0.000,0.000},opacity=0.50} (10.228,0.985)--(10.308,0.985)--(10.308,1.994)--(10.228,1.994)--cycle;
\draw[gp path] (10.228,0.985)--(10.228,1.993)--(10.307,1.993)--(10.307,0.985)--cycle;
\gpfill{rgb color={0.000,0.000,0.000},opacity=0.50} (10.307,0.985)--(10.388,0.985)--(10.388,2.132)--(10.307,2.132)--cycle;
\draw[gp path] (10.307,0.985)--(10.307,2.131)--(10.387,2.131)--(10.387,0.985)--cycle;
\gpfill{rgb color={0.000,0.000,0.000},opacity=0.50} (10.387,0.985)--(10.467,0.985)--(10.467,2.296)--(10.387,2.296)--cycle;
\draw[gp path] (10.387,0.985)--(10.387,2.295)--(10.466,2.295)--(10.466,0.985)--cycle;
\gpfill{rgb color={0.000,0.000,0.000},opacity=0.50} (10.466,0.985)--(10.547,0.985)--(10.547,2.529)--(10.466,2.529)--cycle;
\draw[gp path] (10.466,0.985)--(10.466,2.528)--(10.546,2.528)--(10.546,0.985)--cycle;
\gpfill{rgb color={0.000,0.000,0.000},opacity=0.50} (10.546,0.985)--(10.626,0.985)--(10.626,2.852)--(10.546,2.852)--cycle;
\draw[gp path] (10.546,0.985)--(10.546,2.851)--(10.625,2.851)--(10.625,0.985)--cycle;
\gpfill{rgb color={0.000,0.000,0.000},opacity=0.50} (10.625,0.985)--(10.706,0.985)--(10.706,3.184)--(10.625,3.184)--cycle;
\draw[gp path] (10.625,0.985)--(10.625,3.183)--(10.705,3.183)--(10.705,0.985)--cycle;
\gpfill{rgb color={0.000,0.000,0.000},opacity=0.50} (10.705,0.985)--(10.785,0.985)--(10.785,3.386)--(10.705,3.386)--cycle;
\draw[gp path] (10.705,0.985)--(10.705,3.385)--(10.784,3.385)--(10.784,0.985)--cycle;
\gpfill{rgb color={0.000,0.000,0.000},opacity=0.50} (10.784,0.985)--(10.865,0.985)--(10.865,3.338)--(10.784,3.338)--cycle;
\draw[gp path] (10.784,0.985)--(10.784,3.337)--(10.864,3.337)--(10.864,0.985)--cycle;
\gpfill{rgb color={0.000,0.000,0.000},opacity=0.50} (10.864,0.985)--(10.944,0.985)--(10.944,3.043)--(10.864,3.043)--cycle;
\draw[gp path] (10.864,0.985)--(10.864,3.042)--(10.943,3.042)--(10.943,0.985)--cycle;
\gpfill{rgb color={0.000,0.000,0.000},opacity=0.50} (10.943,0.985)--(11.024,0.985)--(11.024,2.597)--(10.943,2.597)--cycle;
\draw[gp path] (10.943,0.985)--(10.943,2.596)--(11.023,2.596)--(11.023,0.985)--cycle;
\gpfill{rgb color={0.000,0.000,0.000},opacity=0.50} (11.023,0.985)--(11.103,0.985)--(11.103,2.135)--(11.023,2.135)--cycle;
\draw[gp path] (11.023,0.985)--(11.023,2.134)--(11.102,2.134)--(11.102,0.985)--cycle;
\gpfill{rgb color={0.000,0.000,0.000},opacity=0.50} (11.102,0.985)--(11.183,0.985)--(11.183,1.749)--(11.102,1.749)--cycle;
\draw[gp path] (11.102,0.985)--(11.102,1.748)--(11.182,1.748)--(11.182,0.985)--cycle;
\gpfill{rgb color={0.000,0.000,0.000},opacity=0.50} (11.182,0.985)--(11.262,0.985)--(11.262,1.474)--(11.182,1.474)--cycle;
\draw[gp path] (11.182,0.985)--(11.182,1.473)--(11.261,1.473)--(11.261,0.985)--cycle;
\gpfill{rgb color={0.000,0.000,0.000},opacity=0.50} (11.261,0.985)--(11.342,0.985)--(11.342,1.300)--(11.261,1.300)--cycle;
\draw[gp path] (11.261,0.985)--(11.261,1.299)--(11.341,1.299)--(11.341,0.985)--cycle;
\gpfill{rgb color={0.000,0.000,0.000},opacity=0.50} (11.341,0.985)--(11.421,0.985)--(11.421,1.190)--(11.341,1.190)--cycle;
\draw[gp path] (11.341,0.985)--(11.341,1.189)--(11.420,1.189)--(11.420,0.985)--cycle;
\gpfill{rgb color={0.000,0.000,0.000},opacity=0.50} (11.420,0.985)--(11.501,0.985)--(11.501,1.124)--(11.420,1.124)--cycle;
\draw[gp path] (11.420,0.985)--(11.420,1.123)--(11.500,1.123)--(11.500,0.985)--cycle;
\gpfill{rgb color={0.000,0.000,0.000},opacity=0.50} (11.500,0.985)--(11.580,0.985)--(11.580,1.080)--(11.500,1.080)--cycle;
\draw[gp path] (11.500,0.985)--(11.500,1.079)--(11.579,1.079)--(11.579,0.985)--cycle;
\gpfill{rgb color={0.000,0.000,0.000},opacity=0.50} (11.579,0.985)--(11.660,0.985)--(11.660,1.052)--(11.579,1.052)--cycle;
\draw[gp path] (11.579,0.985)--(11.579,1.051)--(11.659,1.051)--(11.659,0.985)--cycle;
\gpfill{rgb color={0.000,0.000,0.000},opacity=0.50} (11.659,0.985)--(11.739,0.985)--(11.739,1.033)--(11.659,1.033)--cycle;
\draw[gp path] (11.659,0.985)--(11.659,1.032)--(11.738,1.032)--(11.738,0.985)--cycle;
\gpfill{rgb color={0.000,0.000,0.000},opacity=0.50} (11.738,0.985)--(11.819,0.985)--(11.819,1.023)--(11.738,1.023)--cycle;
\draw[gp path] (11.738,0.985)--(11.738,1.022)--(11.818,1.022)--(11.818,0.985)--cycle;
\gpfill{rgb color={0.000,0.000,0.000},opacity=0.50} (11.818,0.985)--(11.898,0.985)--(11.898,1.015)--(11.818,1.015)--cycle;
\draw[gp path] (11.818,0.985)--(11.818,1.014)--(11.897,1.014)--(11.897,0.985)--cycle;
\gpfill{rgb color={0.000,0.000,0.000},opacity=0.50} (11.897,0.985)--(11.978,0.985)--(11.978,1.009)--(11.897,1.009)--cycle;
\draw[gp path] (11.897,0.985)--(11.897,1.008)--(11.977,1.008)--(11.977,0.985)--cycle;
\gpfill{rgb color={0.000,0.000,0.000},opacity=0.50} (11.977,0.985)--(12.057,0.985)--(12.057,1.003)--(11.977,1.003)--cycle;
\draw[gp path] (11.977,0.985)--(11.977,1.002)--(12.056,1.002)--(12.056,0.985)--cycle;
\gpfill{rgb color={0.000,0.000,0.000},opacity=0.50} (12.056,0.985)--(12.137,0.985)--(12.137,0.999)--(12.056,0.999)--cycle;
\draw[gp path] (12.056,0.985)--(12.056,0.998)--(12.136,0.998)--(12.136,0.985)--cycle;
\gpfill{rgb color={0.000,0.000,0.000},opacity=0.50} (12.136,0.985)--(12.216,0.985)--(12.216,0.995)--(12.136,0.995)--cycle;
\draw[gp path] (12.136,0.985)--(12.136,0.994)--(12.215,0.994)--(12.215,0.985)--cycle;
\gpfill{rgb color={0.000,0.000,0.000},opacity=0.50} (12.215,0.985)--(12.296,0.985)--(12.296,0.992)--(12.215,0.992)--cycle;
\draw[gp path] (12.215,0.985)--(12.215,0.991)--(12.295,0.991)--(12.295,0.985)--cycle;
\gpfill{rgb color={0.000,0.000,0.000},opacity=0.50} (12.295,0.985)--(12.375,0.985)--(12.375,0.990)--(12.295,0.990)--cycle;
\draw[gp path] (12.295,0.985)--(12.295,0.989)--(12.374,0.989)--(12.374,0.985)--cycle;
\gpfill{rgb color={0.000,0.000,0.000},opacity=0.50} (12.374,0.985)--(12.454,0.985)--(12.454,0.989)--(12.374,0.989)--cycle;
\draw[gp path] (12.374,0.985)--(12.374,0.988)--(12.453,0.988)--(12.453,0.985)--cycle;
\gpfill{rgb color={0.000,0.000,0.000},opacity=0.50} (12.453,0.985)--(12.534,0.985)--(12.534,0.989)--(12.453,0.989)--cycle;
\draw[gp path] (12.453,0.985)--(12.453,0.988)--(12.533,0.988)--(12.533,0.985)--cycle;
\gpfill{rgb color={0.000,0.000,0.000},opacity=0.50} (12.533,0.985)--(12.613,0.985)--(12.613,0.988)--(12.533,0.988)--cycle;
\draw[gp path] (12.533,0.985)--(12.533,0.987)--(12.612,0.987)--(12.612,0.985)--cycle;
\gpfill{rgb color={0.000,0.000,0.000},opacity=0.50} (12.612,0.985)--(12.693,0.985)--(12.693,0.988)--(12.612,0.988)--cycle;
\draw[gp path] (12.612,0.985)--(12.612,0.987)--(12.692,0.987)--(12.692,0.985)--cycle;
\gpfill{rgb color={0.000,0.000,0.000},opacity=0.50} (12.692,0.985)--(12.772,0.985)--(12.772,0.987)--(12.692,0.987)--cycle;
\draw[gp path] (12.692,0.985)--(12.692,0.986)--(12.771,0.986)--(12.771,0.985)--cycle;
\gpfill{rgb color={0.000,0.000,0.000},opacity=0.50} (12.771,0.985)--(12.852,0.985)--(12.852,0.987)--(12.771,0.987)--cycle;
\draw[gp path] (12.771,0.985)--(12.771,0.986)--(12.851,0.986)--(12.851,0.985)--cycle;
\gpfill{rgb color={0.000,0.000,0.000},opacity=0.50} (12.851,0.985)--(12.931,0.985)--(12.931,0.987)--(12.851,0.987)--cycle;
\draw[gp path] (12.851,0.985)--(12.851,0.986)--(12.930,0.986)--(12.930,0.985)--cycle;
\gpfill{rgb color={0.000,0.000,0.000},opacity=0.50} (12.930,0.985)--(13.011,0.985)--(13.011,0.986)--(12.930,0.986)--cycle;
\draw[gp path] (12.930,0.985)--(13.010,0.985)--cycle;
\gpfill{rgb color={0.000,0.000,0.000},opacity=0.50} (13.010,0.985)--(13.090,0.985)--(13.090,0.986)--(13.010,0.986)--cycle;
\draw[gp path] (13.010,0.985)--(13.089,0.985)--cycle;
\gpfill{rgb color={0.000,0.000,0.000},opacity=0.50} (13.089,0.985)--(13.170,0.985)--(13.170,0.986)--(13.089,0.986)--cycle;
\draw[gp path] (13.089,0.985)--(13.169,0.985)--cycle;
\gpfill{rgb color={0.000,0.000,0.000},opacity=0.50} (13.169,0.985)--(13.249,0.985)--(13.249,0.986)--(13.169,0.986)--cycle;
\draw[gp path] (13.169,0.985)--(13.248,0.985)--cycle;
\gpfill{rgb color={0.000,0.000,0.000},opacity=0.50} (13.248,0.985)--(13.329,0.985)--(13.329,0.986)--(13.248,0.986)--cycle;
\draw[gp path] (13.248,0.985)--(13.328,0.985)--cycle;
\gpfill{rgb color={0.000,0.000,0.000},opacity=0.50} (13.328,0.985)--(13.408,0.985)--(13.408,0.986)--(13.328,0.986)--cycle;
\draw[gp path] (13.328,0.985)--(13.407,0.985)--cycle;
\gpfill{rgb color={0.000,0.000,0.000},opacity=0.50} (13.407,0.985)--(13.448,0.985)--(13.448,0.986)--(13.407,0.986)--cycle;
\draw[gp path] (13.407,0.985)--(13.447,0.985)--cycle;
\gpcolor{color=gp lt color border}
\gpsetlinetype{gp lt border}
\draw[gp path] (2.240,6.631)--(2.240,0.985)--(13.447,0.985)--(13.447,6.631)--cycle;
\gpdefrectangularnode{gp plot 1}{\pgfpoint{2.240cm}{0.985cm}}{\pgfpoint{13.447cm}{6.631cm}}
\end{tikzpicture}